\documentclass[iop]{emulateapj}

\slugcomment{accepted by \apj}

\newcommand{\G}{\Gamma}

\newcommand{\sT}{\sigma_{\rm T}}
\newcommand{\p}{^\prime}
\newcommand{\e}{\epsilon}
\newcommand{\g}{\gamma}

\newcommand{\dD}{\delta_{\rm D}}
\newcommand{\up}{u^\prime}

\newcommand{\psim}{\lower.5ex\hbox{$\; \buildrel \propto \over\sim \;$}}
\newcommand{\lbar}{\lower.0ex\hbox{$\; \buildrel
{\lower0.0ex \hbox{-}} \over\lambda  \;$}}

\newcommand{\cm}{\mathrm{cm}}
\newcommand{\km}{\mathrm{km}}
\newcommand{\erg}{\mathrm{erg}}

\newcommand{\s}{\mathrm{s}}

\newcommand{\Hz}{\mathrm{Hz}}
\newcommand{\pc}{\mathrm{pc}}

\newcommand{\Mpc}{\mathrm{Mpc}}

\newcommand{\Kelvin}{\mathrm{K}}

\shorttitle{Blazar Sequence}
\shortauthors{Finke}

\begin{document}
\title{Compton Dominance and the Blazar Sequence}

\author{Justin D.\ Finke}

\affil{U.S.\ Naval Research Laboratory, Code 7653, 4555 Overlook Ave.\ SW,
        Washington, DC,
        20375-5352\\
}

\email{justin.finke@nrl.navy.mil}

\begin{abstract}

Does the ``blazar sequence'' exist, or is it a result of a selection
effect, due to the difficulty in measuring the redshifts of blazars
with both high synchrotron peak frequencies ($\ga 10^{15}\ \Hz$) and
luminosities ($\ga 10^{46}\ \erg\ \s^{-1}$)?  We explore this question
with a sample of blazars from the Second Catalog of Active Galactic
Nuclei (AGN) from the {\em Fermi}\ Large Area Telescope (LAT).  The
Compton dominance, the ratio of the peak of the Compton to the
synchrotron peak luminosities, is essentially a redshift-independent
quantity, and thus crucial to answering this question.  We find that a
correlation exists between Compton dominance and the peak frequency of
the synchrotron component for all blazars in the sample, including
ones with unknown redshift.  We then construct a simple model to
explain the blazar properties in our sample, where the difference
between sources is due to only the magnetic field of the blazar jet
emitting region, the external radiation field energy density, and the
jet angle to the line of sight, with the magnetic field strength and
external energy density being correlated.  This model can reproduce
the trends of the blazars in the sample, and predicts blazars may be
discovered in the future with high synchrotron peak frequencies and
luminosities.  At the same time the simple model reproduces the lack
of high-synchrotron peaked blazars with high Compton dominances ($\ga
1$).

\end{abstract}

\keywords{galaxies: active --- BL Lacertae objects: general ---
quasars: general --- gamma rays: galaxies --- radiation mechanisms:
nonthermal}

\section{Introduction}
\label{intro}

AGN with relativistic jets pointed along our line of sight are known
collectively as blazars \citep[e.g.,][]{urry95}.  Blazars have two
main sub-classes, those with strong broad emission lines, Flat
Spectrum Radio Quasars (FSRQs), and those with weak or absent lines,
known as BL Lacertae objects \citep[BL Lacs;][]{marcha96,landt04}.
FSRQs are thought to be Fanaroff-Riley \citep[FR;][]{fanaroff74} type
II radio galaxies aligned along our line of sight, while BL Lac
objects are thought to be aligned FR type I radio galaxies.  In
general, the spectral energy distributions (SEDs) of blazars have two
basic components: a low frequency component, peaking in the optical
through X-rays, from synchrotron emission; and a high frequency
component, peaking in the $\g$ rays, probably originating from Compton
scattering of some seed photon source, either internal (synchrotron
self-Compton or SSC) or external to the jet (external Compton or EC).

Aside from their classifications as FSRQs or BL Lacs from optical
spectra, \citet{abdo10_sed} subdivided them based on their synchrotron
peak.  They are considered high synchrotron-peaked (HSP) blazars if
their synchrotron peak $\nu^{sy}_{pk} > 10^{15}\ \Hz$; intermediate
synchrotron-peaked (ISP) blazars if $10^{14}\ \Hz < \nu^{sy}_{pk} <
10^{15}\ \Hz$; and low synchrotron-peaked (LSP) blazars if
$\nu^{sy}_{pk}<10^{14}\ \Hz$.  Almost all FSRQs are LSP blazars.

\citet{fossati98} combined several blazar surveys and noted an
anti-correlation between the luminosity at the synchrotron peak,
$L_{pk}^{sy}$, and the frequency of this peak, $\nu^{sy}_{pk}$.  They
also noticed anti-correlations between the 5 GHz luminosity ($L_{5\
GHz}$) and $\nu^{sy}_{pk}$; the $\g$-ray luminosity and
$\nu^{sy}_{pk}$; and the $\g$-ray dominance (the ratio of the EGRET
$\g$-ray luminosity to the synchrotron peak luminosity) and
$\nu^{sy}_{pk}$.  These correlations were claimed as evidence for a
``blazar sequence'', a systematic trend from luminous, low-peaked,
$\g$-ray dominant sources with strong broad emission lines to less
luminous, high-peaked sources with weak or nonexistent broad emission
lines and $\g$-ray dominance $\sim 1$.  \citet{ghisellini98} provided
a physical explanation for these correlations.  If the seed photon
source for external Compton scattering is the broad-line region (BLR),
and the BLR strength is correlated with the power injected into
electrons in the jet, one would expect that more luminous jets have
stronger broad emission lines and greater Compton cooling, and thus a
lower $\nu^{sy}_{pk}$.  As the power injected in electrons is reduced,
the broad line luminosity decreases, there are fewer seed photons for
Compton scattering, and consequently the peak synchrotron frequency
moves to higher frequencies.  This trend is also reflected in the
lower luminosity of the Compton-scattered component relative to the
synchrotron component as $\nu^{sy}_{pk}$ moves to higher frequencies.
If this physical explanation is correct, it provides a powerful
tool for understanding blazars and their evolution, not unlike the
``main sequence'' for stars on the Hertzsprung- Russel diagram.  The
blazar sequence also has implications for ``feedback'', a relationship
where the AGN jet heats the hot, X-ray emitting intracluster medium
(ICM), while the ICM gas Bondi accretes onto the black hole, providing
the fuel for the jet \citep[e.g.,][]{birzan04}.  \citet{hardcastle07}
have suggested that radio galaxies without high excitation narrow
lines (mostly FR Is, and presumably BL Lacs) are fed by ``hot mode''
accretion, i.e., accretion from the hot ICM; while radio galaxies with
high excitation narrow lines (mostly FR IIs, and presumably FSRQs) are
fed by ``cold mode'' accretion, i.e., accretion of cold gas unrelated
to the ICM.

The $L_{pk}^{sy}$--$\nu_{pk}^{sy}$ anti-correlation has been
questioned.  Using blazars from the Deep X-ray Radio Blazar Survey and
the ROSAT All-Sky Survey/Green Bank Survey, \citet{padovani03} did not
find any anti-correlation between $\nu^{sy}_{pk}$ and radio, BLR, or
jet power.  This work, however, has been criticized for its relatively
poor SED characterization \citep{ghisellini08_seq}.
\citet{padovani07} has identified several major predictions of the
blazar sequence: the anti-correlation will continue to be found
in more complete samples; since low luminosity objects are almost
always more plentiful than high luminosity objects, high-peaked
blazars should be more plentiful than low-peaked blazars; and the lack
of plentiful outliers, i.e., objects that are low-peaked and
faint, or high-peaked and bright. If any of these predictions are
contradicted, it would invalidate the sequence.

\citet{ghisellini08_seq} pointed out that since blazars are
anisotropic emitters, the lower left of the diagram (i.e.,
sources that have low peaked, faint synchrotron components) should be
filled in by sources which are viewed increasingly off-axis.  But the
lack of sources in the upper right region of the
$L_{pk}^{sy}$--$\nu_{pk}^{sy}$ plot (i.e., sources that have
high-peaked, bright synchrotron components) could be the result of a
selection effect
\citep{giommi02,padovani02,giommi05,giommi12_selection}.  Since a
large fraction of BL Lac objects have entirely featureless optical
spectra, their redshifts, $z$, and hence luminosities, are impossible
to determine.  These could be extremely luminous, distant BL Lacs that
would fill in the upper right region.  \citet{nieppola06} found no
correlation between the frequency and luminosity of the synchrotron
peaks for objects in the Mets\"ahovi Radio Observatory BL Lacertae
sample.  \citet{chen11} did find an anti-correlation, using sources
found in the LAT bright AGN sample \citep{abdo09_lbas,abdo10_sed}.  In
recent works, an ``L''-shape in the $L_{pk}^{sy}$--$\nu_{pk}^{sy}$
plot seems to have emerged, as lower luminosity low-peaked sources
have been detected with more sensitive instruments
\citep{meyer11,giommi12}.  \citet{nieppola06} found more of a ``V''
shape, although their plot did not include FSRQs; if high luminosity
and low-peaked FSRQs were added, it might appear as more of an ``L''.
None of the studies that find few sources with high $L_{pk}^{sy}$ and
high $\nu_{pk}^{sy}$ have yet to address the sources without
redshifts, however.  \citet{rau12} provide reliable photometric
redshifts for 8 BL Lacs at $z\ga1.3$, and four of these do indeed seem
to have $\nu_{pk}^{sy}\ga10^{15}$\ Hz and $L_{pk}^{sy}\ga 10^{46}\
\erg\ \s^{-1}$ \citep{padovani12}.

With the advent of the {\em Fermi Gamma-Ray Space Telescope} era, it
is now possible to also characterize the Compton peak frequency
$\nu_{pk}^C$ and luminosity $L_{pk}^C$ for a larger number of
objects than previously possible with EGRET.  As we show, the Compton
dominance $A_C\equiv L_{pk}^{C} / L_{pk}^{sy}$ is an important
parameter for characterizing this sequence, since it is a
redshift-independent quantity.  We explore the sequence using a sample
based on the second catalog of AGN from the LAT
\citep[2LAC;][]{ackermann11_2lac} in Section \ref{2lac_sequence}.  We
then show in Section \ref{theory} that the sequence, including sources
with high $L_{pk}^{sy}$ and high $\nu_{pk}^{sy}$ can be reproduced
with a simple model involving power-law injection of electrons and
radiative cooling.  Finally, we conclude with a discussion of these
results (Section \ref{discussion}).

\section{The 2LAC Blazar Sequence}
\label{2lac_sequence}

\subsection{Sample Definition and SED Characterization}
\label{sample_section}

\begin{figure}
\vspace{5.0mm}
\epsscale{1.0}
\plotone{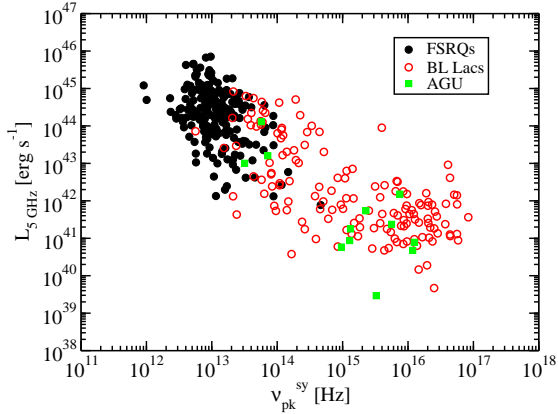}
\caption{Radio luminosity at 5~GHz versus peak synchrotron frequency for 
the 2LAC clean sample.  Symbols are the same as in Figure \ref{fossati}.  
}
\label{fossati_radio}
\vspace{2.2mm}
\end{figure}

The 2LAC \citep{ackermann11_2lac} presents the largest $\g$-ray
catalog of blazars yet.  It allows for the characterization of the
high energy component for a greater number of blazars than previously
possible.  Here we look at the blazar sequence among the 2LAC clean
sample, which includes 886 total sources, with 395 BL Lacs, 310 FSRQs,
and 157 sources of unknown type.  \citet{ackermann11_2lac} collected
the fluxes of many of these sources at 5 GHz, 5000~\AA, and 1 keV, and
used empirical relations for finding the peak frequency of the
synchrotron component from the slope between the 5 GHz and 5000~\AA\
flux ($\alpha_{ro}$), and between the 5000~\AA\ and 1 keV flux
($\alpha_{ox}$) found by \citet{abdo10_sed}.  \citet{abdo10_sed} in
turn had found these empirical relations by fitting the broadband SEDs
of 48 of the blazars in the 3-month LAT bright AGN sample
\citep[LBAS;][]{abdo09_lbas} with third degree polynomials to
determine the peak synchrotron frequency, $\nu_{pk}^{sy}$.  With the
$\nu_{pk}^{sy}$ estimated from the empirical relations,
\citet{ackermann11_2lac} could classify the objects in the 2LAC sample
as LSP, ISP, or HSP.  Here we use their results for $\nu_{sy}^{pk}$,
which are corrected for redshift, and so are in the sources' rest
frames.  The 2LAC also includes the 5 GHz flux density for many
sources.  We combine this with a spectroscopic redshift measurement,
if available, or photometric redshifts found by \citet{rau12}, to get
the luminosity distance, $d_L$\footnote{To calculate $d_L$, we use a
flat $\Lambda$CDM cosmology with $H_0=71\ \km\ \s^{-1}\ \Mpc^{-1}$,
$\Omega_m=0.27$, and $\Omega_\Lambda = 0.73$.} and calculate the 5 GHz
luminosity $L_{5\ GHz}=4\pi d_L^2 (\nu F_{\nu})_{5\ GHz}$.  We have
corrected the radio luminosities to how they would appear in the rest
frames of the sources (i.e., $k$-corrected them), assuming
$\alpha_r=0$, where the radio flux density is
$F_\nu\propto\nu^{-\alpha_r}$.  A plot of $L_{5\ GHz}$ versus
$\nu_{pk}^{sy}$ for the objects in the 2LAC clean sample for which the
catalog has a listed $\nu^{sy}_{pk}$, 5 GHz flux density, and $z$
\citep[or a $z$ from][]{rau12} is given in Figure \ref{fossati_radio}.
This includes 352 sources, of which 145 are BL Lacs, 195 are FSRQs,
and 12 are AGN of unknown optical spectral type (AGUs; i.e., unknown
whether they are FSRQs or BL Lacs).  

\begin{figure}
\vspace{2.2mm} \
epsscale{1.0} 
\plotone{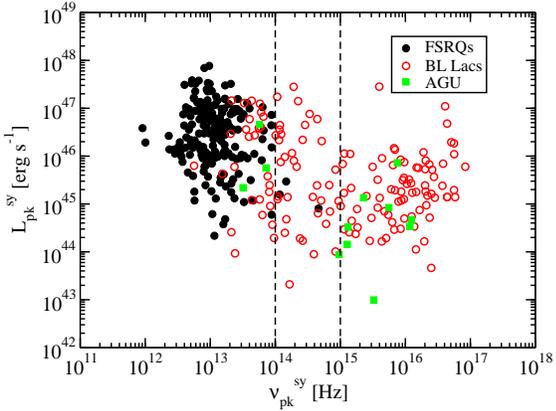}
\caption{Peak synchrotron luminosity versus peak synchrotron frequency
for objects in the 2LAC clean sample.  Filled circles represent FSRQs,
empty circles represent BL Lacs, and filled squares represent objects
which do not have an unambiguous classification.  Dashed lines indicate 
the boundary between HSPs and ISPs ($\nu_{pk}^{sy}=10^{15}\ \Hz$) 
and between ISPs and LSPs ($\nu_{pk}^{sy}=10^{14}\ \Hz$).  
}
\label{fossati}
\vspace{2.2mm}
\end{figure}

\citet{abdo10_sed} also provided an empirical formula for determining
the flux at the synchrotron peak ($F_{pk}^{sy}$) from the the 5 GHz
flux density and $\nu_{pk}^{sy}$.  We use this empirical relation to
determine the peak synchrotron flux for the 352 objects in our
sample, and again combine it with $d_L$ to calculate the peak
synchrotron luminosity, $L_{pk}^{sy}$.  A resulting plot of
$L_{pk}^{sy}$ versus $\nu_{pk}^{sy}$ is shown in Figure
\ref{fossati}.

\begin{figure}
\vspace{6.5mm}
\epsscale{1.0}
\plotone{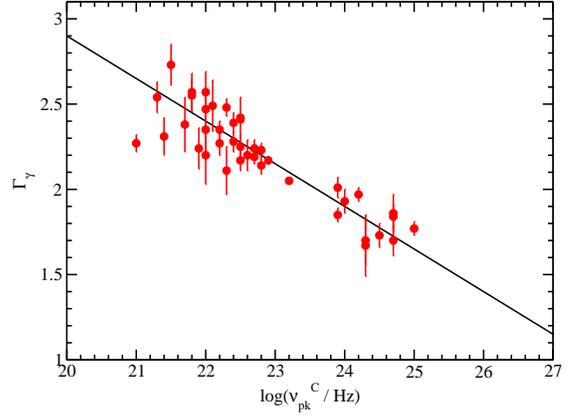}
\caption{The LAT spectral index ($\G_{\g}$) and the peak 
frequency of the Compton component, $\nu_{pk}^C$ determined 
from third-order polynomial fits from \citet{abdo10_sed}, plotted as 
circles.  The empirical fit they determined is plotted as the line.
}
\label{peak_Gamma}
\vspace{2.2mm}
\end{figure}

Additionally, \citet{abdo10_sed} fit the high energy components of
their 48 LBAS blazars with a third degree polynomial to
determine the peak of the $\g$-ray component (presumably from Compton
scattering).  They found an empirical relation between $\nu_{pk}^{C}$
and the LAT $\g$-ray spectral index, $\G_\g$.  Figure \ref{peak_Gamma}
shows $\G_\g$ and $\nu_{pk}^C$ from their polynomial fits to the
objects in their sample, and the empirical relation they found.  The
empirical fit seems reasonable in the range $1.6<\G_\g < 2.6$, but
extending the relation outside this range is questionable.
Approximately 10\% of the 352 sources are also found in the 58-month
{\em Swift} Burst Alert Telescope (BAT)
catalog\footnote{\url{http://heasarc.nasa.gov/docs/swift/results/bs58mon/}}.
For these sources, we extrapolated their BAT and LAT power-laws and
found where the extrapolations intersected.  If they intersected
within the range 195 keV to 100 MeV (that is, in between the BAT
and LAT bandpasses) we used this location as $\nu_{pk}^C$.  Using the
BAT spectrum in this way allows an improved estimation of $\nu_{pk}^C$
over the empirical relation from \citet{abdo10_sed}, particularly for
those very soft sources, for which this empirical relation is
untested.  For other sources, the peak was determined from the LAT
spectral index and the empirical relation from \citet{abdo10_sed}.  In
principle, a similar technique could be used by combining the LAT
spectra and very-high energy (VHE) spectra taken from atmospheric
Cherenkov telescopes, in order to find $\nu_{pk}^C$ for hard LAT
sources.  However, the VHE spectra tend to be highly variable, and are
not integrated over a long period of time, and often are not
simultaneous with the LAT spectra.  Since the BAT and LAT spectra
overlap in time and are integrated over long times, it seems more
reasonable that they would represent a similar state.

\begin{figure}
\vspace{6.0mm}
\epsscale{1.0}
\plotone{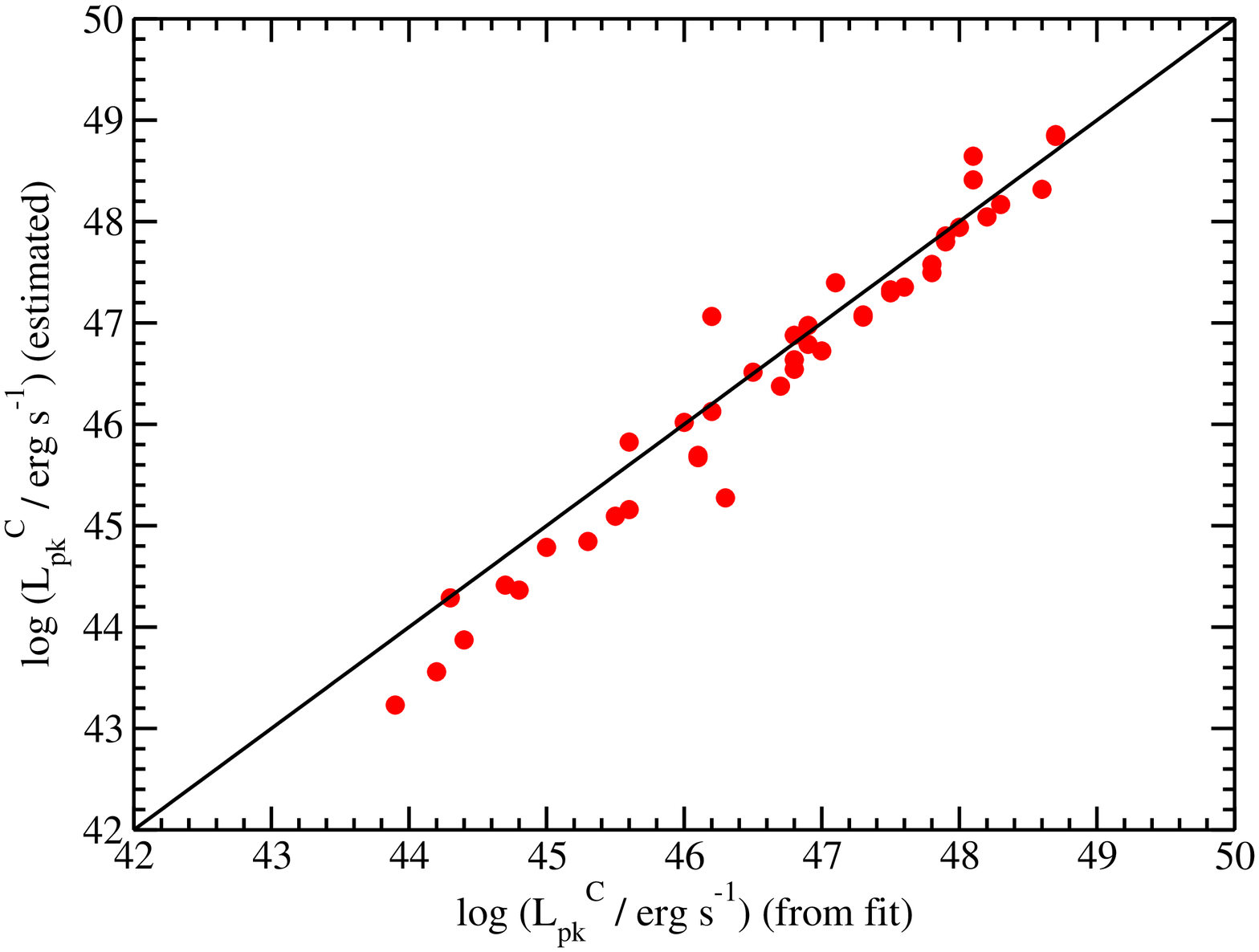}
\caption{The peak Compton luminosity, $L_{pk}^C$, estimated from 
empirical relations as
discussed in the text, plotted versus $L_{pk}^C$ determined from the
fits by \citet{abdo10_sed}.  The line shows where the estimated
and fit values would be equal.  }
\label{peak_L_C}
\vspace{2.2mm}
\end{figure}

\begin{figure}
\vspace{4.0mm}
\epsscale{1.0}
\plotone{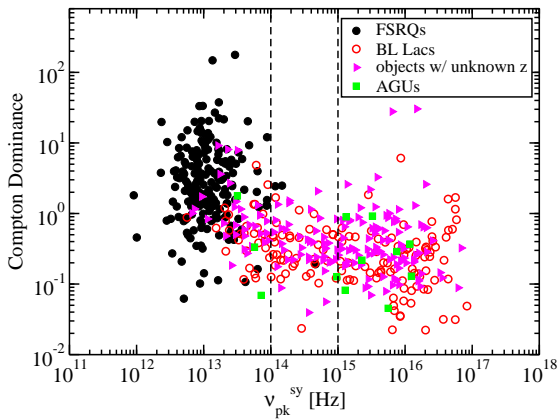}
\caption{Compton dominance (i.e.,
$L_{pk}^C/L_{pk}^{sy}$)  versus peak synchrotron
frequency.  Symbols are the same as in Figures \ref{fossati_radio} and
\ref{fossati}; additionally, rightward-pointing triangles
represent BL Lacs with unknown redshifts, for which $\nu_{pk}^{sy}$ is
a lower limit.  }
\label{CD}
\end{figure}

Once the location of $\nu_{pk}^C$ is known, either by using the
empirical relation or from the BAT-LAT intersection, the flux at the
peak, $F_{pk}^{C}$, can be estimated by extrapolating the LAT
spectrum, and the corresponding luminosity can be calculated with
$L_{pk}^C = 4\pi d_L^2 F_{pk}^{C}$.  To test the accuracy of this
approximation, we use this method to estimate $L_{pk}^C$ for the
sample in \citet{abdo10_sed}, and compare it with the value found by
their fits to 48 sources.  The results can be found in Figure
\ref{peak_L_C}; the agreement seems reasonable.  Therefore, we used
this technique to estimate $L_{pk}^C$ for the 352 objects in our
sample.  With this, we can calculate the Compton dominance, $A_C =
L_{pk}^{C} / L_{pk}^{sy}$.  A plot of $A_C$ versus
$\nu_{pk}^{sy}$ is shown in Figure \ref{CD}.  Also note that $A_C$ is
independent of redshift ($A_C = L_{pk}^{C} / L_{pk}^{sy} \approx
F_{pk}^C / F_{pk}^{sy}$).  Thus we can plot in Figure \ref{CD} an
additional 170 sources from the 2LAC clean sample that have
well-determined synchrotron bumps but do not have known redshifts.
For these sources, the plotted $\nu_{pk}^{sy}$ is a lower limit, since
the redshifts are not known.  However, $\nu_{pk}^{sy}$ will be larger
by only a factor $(1+z)$, i.e., a factor of a few.

\vspace{2.2mm}

\subsection{Luminosity and the Sequence}

In their diagrams of $L_{pk}^{sy}$ versus $\nu_{pk}^{sy}$,
\citet{nieppola06} and \citet{meyer11} note a ``V'' or ``L'' shape.
This shape is not at all obvious in our version of this plot in Figure
\ref{fossati}.  To determine the strength of the correlation between
$L_{pk}^{sy}$ and $\nu_{pk}^{sy}$, we have computed the Spearman
($\rho$) and Kendall ($\tau$) rank correlation coefficients and the
probability of no correlation (PNC) calculated from each coefficient.
The results can be found in Table \ref{table_correlate}.  The results
from $\rho$ and $\tau$ are similar in all cases.  The PNC is very
small for the BL Lacs and FSRQs separately, and even lower for all the
sources combined, where the probability is essentially zero that there
is not a correlation.  The small PNC is consistent with
\citet{fossati98}, who find essentially a zero chance of no
correlation using the $\tau$ test.  Note however that sources with
unknown $z$ are not included.  Since these sources could have high
$L_{pk}^{sy}$, their exclusion could explain the anti-correlations for
the whole sample and for the BL Lacs in general
\citep{giommi12_selection}, since they could fill in the upper right
part of this diagram, as mentioned in Section \ref{intro}.  Indeed,
several sources with redshifts estimated from photometry have been
found in this region \citep{padovani12,rau12}, and are included in our
sample.  The objects with unknown $z$ are almost certainly BL Lacs
(i.e., they probably have small broad-line equivalent widths, making
their redshifts difficult to determine), so it would not affect the
hint (92.7\% and 93.4\% for the $\rho$ and $\tau$ coefficients,
respectively) of anti-correlation found for FSRQs.  It should also be
noted that \citet{nieppola06} did not find a correlation between
$L^{sy}_{pk}$ and $\nu^{sy}_{pk}$ for BL Lacs using the $\rho$
coefficient, possibly because they used log-parabola functions to
characterize the SEDs, while \citet{fossati98} and \citet{abdo10_sed}
use third-degree polynomials.  Since our determinations of
$\nu_{pk}^{sy}$ and $L_{pk}^{sy}$ are based on empirical relations
related to these third-degree polynomial fits, our results could also
be expected to be different from those of \citet{nieppola06}.

\citet{fossati98} also found a significant anti-correlation between
the 5~GHz luminosity, $L_{5\ GHz}$, and $\nu_{pk}^{sy}$, using the
$\tau$ test.  Like them, we find the anti-correlation with $L_{5\
GHz}$ versus $\nu_{pk}^{sy}$ (Figure \ref{fossati_radio}) to be much
more significant than the one between $L_{pk}^{sy}$ and
$\nu_{pk}^{sy}$, as can be seen in Table \ref{table_correlate}.  This
is true for both FSRQs and BL Lacs alone as well as together.  If the
anti-correlation is explained by the increasing cooling at higher
luminosities \citep{ghisellini98}, one might expect the correlation
with $L^{sy}_{pk}$ to be more significant, since the emission at 5 GHz
is thought to be from a different region of the jet than the emission
at the peak, and the region emitting at the peak should be synchrotron
self-absorbed at 5 GHz.  However, since $\nu_{pk}^{sy}$ was determined
in part based on the 5 GHz flux, the strong correlation with $L_{5\
GHz}$ is probably a result of this dependence.  As pointed out by
\citet{lister11}, if $\nu_{pk}^{sy}$ increases, but its spectral
shape and $\nu F_\nu$ peak flux do not change, the radio flux (or
luminosity) will naturally decrease.  This can easily explain this
anti-correlation.  However, what is unclear is whether a peak
frequency derived from the radio flux should be interpreted as a
cooling break, since these two should be from different regions and
possibly independent.  Determination of $\nu_{pk}^{sy}$ independent of
low radio frequency emission should be a good, although technically
challenging, way to test this.

\begin{deluxetable*}{lcccc}
\tabletypesize{\scriptsize}
\tablecaption{Statistics of correlations 
involving $\nu_{pk}^{sy}$.  
}
\tablewidth{0pt}
\tablehead{
\colhead{Sample} & 
\colhead{$\rho$} & 
\colhead{PNC($\rho$)} & 
\colhead{$\tau$} & 
\colhead{PNC($\tau$)}
}
\startdata
\multicolumn{5}{c}{$L_{pk}^{sy}$ versus $\nu_{pk}^{sy}$}  \\
\hline
BL Lacs & -0.19 & 0.020 & -0.12 & 0.035 \\
FSRQs & -0.12 & 0.073 & -0.088 & 0.066 \\
All sources with known $z$ & -0.54 & $4.2\times10^{-28}$ & -0.35 & $<10^{-50}$ \\
\hline
\multicolumn{5}{c}{$L_{5GHz}$ versus $\nu_{pk}^{sy}$}  \\
\hline
BL Lacs & -0.64 & $5.4\times10^{-18}$ & -0.45 & $<10^{-50}$ \\
FSRQs & -0.36 & $1.82\times10^{-7}$ & -0.25 & $1.79\times10^{-7}$ \\
All sources with known $z$ & -0.79 & $<10^{-50}$ & -0.58 & $<10^{-50}$ \\
\hline
\multicolumn{5}{c}{$A_C$ versus $\nu_{pk}^{sy}$}  \\
\hline

BL Lacs & -0.30 & $2.7\times10^{-4}$  & -0.21 &  $1.3\times10^{-4}$  \\
FSRQs & $8.9\times10^{-3}$ & 0.90 & $6.3\times10^{-3}$ & 0.89 \\
All sources with known $z$ & -0.66 &  $9.8\times10^{-45}$  & -0.45 & $<10^{-50}$ \\
All sources 1\tablenotemark{a} & -0.66 & $<10^{-50}$ &  -0.46 & $<10^{-50}$ \\
All sources 2\tablenotemark{b} & -0.66 & $<10^{-50}$ &  -0.46 & $<10^{-50}$ \\
All sources 3\tablenotemark{c} & -0.64 & $<10^{-50}$ &  -0.44 & $<10^{-50}$ 
\enddata
\tablenotetext{a}{All sources, including those with unknown $z$, assuming they are at $z=0.00$.}
\tablenotetext{b}{All sources, including those with unknown $z$, assuming they are at $z=0.35$.}
\tablenotetext{c}{All sources, including those with unknown $z$, assuming they are at $z=4.00$.} 
\label{table_correlate}
\end{deluxetable*}

\subsection{Compton Dominance and the Sequence}

Combining their results with EGRET data, \citet{fossati98} made a plot
of $\g$-ray dominance versus $\nu_{pk}^{sy}$.  Using LAT data from the
2LAC, as described in Section \ref{sample_section}, we make a
similar plot (Figure \ref{CD}), although we use $L_{pk}^C$
instead of simply the $\g$-ray luminosity.  A distinct ``L'' shape is
seen in this figure.  We have also computed the correlation
coefficients $\rho$ and $\tau$ for $A_C$ versus $\nu_{pk}^{sy}$, and
the results can be found in Table \ref{table_correlate}.  First
we do this only for sources with known $z$.  There is no evidence for
a correlation for the FSRQs alone, although the probability that there
is no correlation for the BL Lacs alone is very small.  For the
combined sample of all sources with known $z$, there is essentially
zero chance that there is no correlation, similar to the $L_{pk}^{sy}$
versus $\nu_{pk}^{sy}$ correlation.  \citet{fossati98} also find a low
probability that there is no correlation using the $\tau$ test,
although our result is stronger.  We also computed the coefficients
for all sources, including the ones with unknown $z$, computing their
$\nu_{pk}^{sy}$ assuming $z=0$ (``all sources 1'' in Table
\ref{table_correlate}).  We find essentially no chance that the
addition of sources with unknown $z$ could ruin the correlation
when these sources are included.  The term $\nu_{pk}^{sy}$ will vary
by a factor of a few due to redshift, so we also calculated the
coefficients assuming all these sources with unknown $z$ are at
$z=0.35$, the average of the BL Lacs with known $z$ (``all sources
2''); and assuming these sources are at $z=4$ (``all sources 3''),
which is significantly higher than the maximum redshift of the entire
sample (which is $z=3.1$).  In each case, there is essentially a 100\%
chance that an anti-correlation exists.  {\em Although the
correlation between $L_{pk}^{sy}$ and $\nu_{pk}^{sy}$ could exist only
because of a redshift selection effect, a similar redshift selection
effect cannot explain the correlation between $A_C$ and
$\nu_{pk}^{sy}$.  The relationship between $A_C$ and $\nu_{pk}^{sy}$
thus seems to have a physical origin.} 

Although a similar quantity, $\g$-ray dominance, was discussed as part
of the original ``blazar sequence'' by \citet{fossati98}, $\g$-ray
dominance or Compton dominance has been mostly overlooked ever since,
in favor of testing the correlation between $L_{pk}^{sy}$ and
$\nu_{pk}^{sy}$
\citep[e.g.,][]{padovani03,nieppola06,padovani07,meyer11,giommi12_selection},
although see \citet{giommi12} for a brief discussion.  This neglect
may be in part due to a lack of quality $\g$-ray data, a deficiency
that has been corrected in the {\em Fermi} era.

\subsection{Errors and Outliers}

The errors on determining $\nu_{pk}^{sy}$ and $L_{pk}^{sy}$ have
essentially been ignored in the past.  Although for radio and optical
observations measurement errors are small, clearly the peak will
depend on how well-sampled the synchrotron bump is.  Furthermore,
X-ray measurements can have large errors (due in part to the
necessity of assuming a spectral form to convolve with the response
matrix of an X-ray instrument), and here ignoring their errors could
lead to significant errors on $\nu_{pk}^{sy}$ and $L_{pk}^{sy}$.  It
is also possible that the peak location can depend on the fitting
function used.  It does not seem to be standardized; sometimes a
log-parabola function is used \citep[e.g.,][]{nieppola06}, sometimes a
third degree polynomial is used
\citep[e.g.,][]{fossati98,abdo10_sed,meyer11} and sometimes a
physically-motivated synchrotron/Compton model is used
\citep[e.g.,][]{padovani03}, although the third-order polynomial fit
is the most common.

When locating the Compton peak, the measurement errors can have an
even greater effect, since the $\g$-ray error bars tend to be larger
than at lower frequencies.  This can lead to significant errors in
determining $A_C$.  \citet{abdo10_sed} have estimated the error
between the approximate expressions and the polynomial fits to be
about an order of magnitude.  This estimate still neglects the
measurement errors, which can have a large effect on results.

The two sources with unknown redshifts in the upper right quadrant of
Figure \ref{CD} are 2FGL~J0059.2-0151 (1RXS~005916.3-015030) and
2FGL~J0912.5+2758 (1RXS~J091211.9+27595) with LAT spectral indices of
$\G_\g=1.15\pm0.36$ and $\G_\g=1.20\pm0.37$, respectively.  These are
the two hardest sources in the 2LAC, and the sources with the largest
error bars on their spectral indices; only one source in the 2LAC
clean sample is fainter than these sources (2FGL~J1023.6+2959).  They
are clearly outliers.  Propagating the error on their spectral
indices, one finds that they have Compton dominances of
$\log_{10}(A_C) = 1.44\pm 1.82$ and $\log_{10}(A_C) = 1.48\pm1.98$,
respectively; they have $A_C$ consistent with unity within their error
bars, and so are consistent with the ``L'' shape seen in Figure
\ref{CD}.

%

There are several FSRQ outliers in Figure \ref{CD} as well;
perhaps most interesting are those with $A_C>10^2$.  These three
sources are: 2FGL~J1017.0+3531 associated with B2~1015+35B;
2FGL~J1154.4+6019 associated with CRATES~J1154+6022; and the most
extreme, 2FGL~J1522.0+4348 associated with B3~1520+437, with
$A_C=1500$, which is not shown in Figure \ref{CD}.  The sources
2FGL~J1017.0+3531 and 2FGL~J1522.0+4348 are estimated to have Compton
components peaking in the BAT bandpass, both with peak fluxes
$>6\times10^{-10}\ \erg\ \s^{-1}\ \cm^{-2}$, and both would then be
visible with BAT.  The source 2FGL~J1154.4+6019 is estimated to have a
high-energy peak observed at 577 keV.  Extrapolating from this peak
back into the BAT bandpass with a spectral index $\G=1.5$ gives a flux
of $9.6\times10^{-11}\ \erg\ \s^{-1}\ \cm^{-2}$ at 100 keV.  There are
several objects in the BAT catalog with fluxes less than this, so the
BAT would probably have detected this source as well, if this estimate
is correct.  Thus, these three sources probably have a low-energy
cutoff in their Compton component before the BAT waveband, and their
$A_C$ shown here are probably not accurate.


Variability is another factor that can affect correlations determined
here.  Blazars are highly variable at all wavelengths, and the peaks
of blazars can vary by several orders of magnitude
\citep{fossati00,costamante01}.  \citet{giommi12} explore the issue of
characterizing peaks at different epochs from microwave, X-ray, and
$\g$-ray data.  They conclude that microwave variations are rather
minimal and make little difference in characterizing SEDs, but
variations in X-rays (0.1 - 2.4 keV) and $\g$-rays (in the LAT energy
band) can vary by up to a factor of 10 or greater.

%

For all of these reasons, the results in Figures \ref{fossati_radio},
\ref{fossati}, and \ref{CD} are probably not accurate for individual
sources, although no one has rigorously estimated their errors.
However, we are more interested in the overall trend than individual
sources, so the trend could still be present even for large errors.
To test if it will still be present, we performed a simple Monte Carlo
simulation.  For each source, we randomly drew the LAT spectral
index ($\G_\g$) and luminosity ($L_\g$) from the measured values and
measurement errors, assuming the errors are described by normal
distributions.  Then we used this randomly-drawn value to calculate
$A_C$.  There are no reported errors on $\nu_{pk}^{sy}$, so we assumed
that $\log(\nu_{pk}^{sy})$ has normally-distributed errors of one
decade.  We then used the Spearman and Kendall tests on the
randomly-drawn distribution to determine if the correlation persists.
We performed this simulation $10^5$ times, and found that a PNC $>
5.7\times10^{-7}$ (the probability corresponding to $5\sigma$,
assuming normally distributed errors) was found in only around 5\% of
the simulations, for both $\rho$ and $\tau$ tests.  Thus, we are
confident that the trend persists despite large (random) errors.
There do not seem to be any systematic errors in using the empirical
relations instead of the polynomial fits \citep[][Figures
\ref{peak_Gamma} and \ref{peak_L_C}]{abdo10_sed}, except possibly a
small systematic underestimation of $L_{pk}^{C}$ as shown in Figure
\ref{peak_L_C}.

\section{Theoretical Blazar Sequence}
\label{theory}

\subsection{Simple Unified Blazar Emission Model}
\label{blazarmodel}

We describe a simple model for blazar jet emission, and show that it
can reproduce blazar properties described in Section
\ref{2lac_sequence}.  This model is similar to the ones presented
by \citet{boett02_seq} and \citet{ghisellini08_seq,ghisellini09}.  
We assume the relativistic jet is dominated by emission from a single
zone which is spherical with radius $R^{\p}_b$ in its comoving frame,
and moving with highly relativistic speed $\beta c$ giving it a
Lorentz factor $\G=(1-\beta^2)^{-1/2}$.  The jet makes an angle to the
line of sight $\theta$ so that the Doppler factor $\dD=[\G(1-\beta
\cos\theta)]^{-1}$.  Electrons are injected with a power-law
distribution given by
\begin{eqnarray}
\label{electronaccel}
Q_e(\g) = Q_0 \g^{-q}\ H(\g;\g_1,\g_2)\ ,
\end{eqnarray}
where the Heaviside function $H(x;a,b)=1$ for $a<x<b$ and $=0$
everywhere else.  In terms of a blast wave model, high $\g_1$ can come
about in highly energetic, fast shocks, where a high fraction of the
swept up energy is used to accelerate particles, and the lower-energy
part of the distribution becomes particle-starved
\citep[e.g.,][]{dermer00,dermer09_book}.  The hard X-ray spectra in
some blazars may indicate very hard electron spectra at lower energies
\citep{sikora09}.  Although blazars can be quite variable on
timescales as short as hours \citep[e.g.,][]{abdo09_3c454.3} or even
minutes \citep{aharonian07_2155}, we will assume their average or
quiescent emission can be described by a steady state solution to the
electron continuity equation, where continuous injection is balanced
by cooling and escape.  The power continuously injected in electrons 
is given by
\begin{eqnarray}
L_{inj,e} = m_e c^2 \int^{\g_2}_{\g_1}\ d\g\ \g\ Q_e(\g)\ ,
\end{eqnarray}
or, using Equation (\ref{electronaccel}), 
\begin{equation}
L_{inj,e} = m_e c^2 Q_0 \left\{ \begin{array}{ll}
  (\g_1^{2-q} - \g_2^{2-q})/(q-2) & q \ne 2 \\
  \ln(\g_2/\g_1) & q = 2
  \end{array}
\right. \ .
\end{equation}
We assume an energy-independent escape
timescale given by
\begin{eqnarray}
t_{esc} = \frac{\eta R_b^\prime}{c}
\end{eqnarray}
where $R_b^\prime$ is the comoving radius of the blob and $\eta$ is a
constant $> 1$.  In this case, where $\g_1 < \g_c$ (the {\em
slow-cooling regime}) the electron distribution can be approximated as
  
\begin{equation}
N_e(\g) \approx Q_0 t_{esc}\g_c^{-q} \left\{ \begin{array}{ll}
  (\g/\g_c)^{-q}   & \g_1 < \g < \g_c \\
  (\g/\g_c)^{-q-1} & \g_c < \g < \g_2 
      \end{array}
\right. \ . 
\end{equation}
If $\g_c < \g_1$, i.e.,  the {\em fast-cooling regime}, 
\begin{equation}
\label{fast_cool_elec}
N_e(\g) \approx Q_0 t_{esc}\g_c\g_1^{-(q+1)}  \left\{ \begin{array}{ll}
  (\g/\g_1)^{-2}   & \g_c < \g < \g_1 \\
  (\g/\g_1)^{-q-1} & \g_1 < \g < \g_2
      \end{array}
\right. \ .
\end{equation}
  
Here we assume the electrons are cooled by synchrotron emission and
Thomson scattering, so that
 
\begin{eqnarray}
\label{gc}
\g_c = \frac{ 3 m_e c^2}{ 4c \sT ( \up_B + \up_{sy,tot} + \G^2u_{ext} ) t_{esc} }\ 
\end{eqnarray}
 
is the cooling electron Lorentz factor, where primes denote quantities
in the comoving frame of the blob.  In this equation, the magnetic field energy density
in the blob comoving frame is
\begin{eqnarray}
\up_B = \frac{B^2}{8\pi}\ ,
\end{eqnarray}
the total synchrotron energy density is
\begin{eqnarray}
\label{upsy}
\up_{sy,tot} = \frac{ \sT \up_B}{\pi R_b^{\p 2}} \ \int d\g\ \g^2\ N_e(\g)\ ,
\end{eqnarray}
\citep[e.g.,][]{boett97} and the external energy density $u_{ext}$ is
assumed to be isotropic in the proper frame of the AGN.  As discussed
in Section \ref{intro}, the exact nature of the external radiation
field is not known, and may not even be the same for all blazars.

The two main differences between our model and that of
\citet{boett02_seq} are (1) we relax the assumption that
$\delta_D=\Gamma$; (2) we do not assume that the magnetic field energy
density is a constant fraction of the electron energy density,
\begin{eqnarray}
\up_e = \frac{m_e c^2}{4\pi R_b^{\p 3}/3} \int d\g\ \g\ N_e(\g)\ .
\end{eqnarray}
Since spectral modeling of blazars finds that the ratio $\up_B/\up_e$
vary by $\approx5$ orders of magnitude from source to source
\citep[e.g.,][]{ghisellini10}, we do not feel this assumption is
well-justified.  We assume that a fraction ($\tau$) of the accretion
disk makes up the external radiation field ($u_{ext}$).  Following
\citet{boett02_seq}, we assume that $\tau \propto L_d$, the accretion
disk luminosity, so that $u_{ext} \propto \tau L_d \propto L_d^2$.
From the Blandford-Znajek mechanism \citep{blandford77}, one finds the
power extracted from the black hole rotation to be $L_{BZ}\propto
B_m^2$, where $B_m$ is the magnetic field in the magnetosphere near
the black hole \citep[e.g.][]{caval02}.  If $L_d \propto L_{BZ}$ and
$B\propto B_m$, where $B$ is the magnetic field in the primary jet
emitting region, then one finds that $u_{ext}\propto B^4$.  Therefore,
we decrease the magnetic field assuming
\begin{eqnarray}
\label{B_uext_eqn}
B= B_0 \left( \frac{u_{ext}}{u_0} \right)^{1/4}\ , 
\end{eqnarray}
consistent with \citet{boett02_seq}.  

In the slow-cooling regime, Equation (\ref{upsy}) can be integrated to give
\begin{eqnarray}
\label{usy_slow}
\up_{sy,tot} = \frac{\sT\up_B Q_0 t_{esc}\g^{-q}}{\pi R^{\p 2}_B}
\biggr[ \g_c^q\frac{ \g_c^{-q+3} - \g_1^{-q+3}}{3-q} + 
\\ \nonumber 
  \g_c^{q+1}\ \frac{ \g_2^{-q+2} - \g_c^{-q+2} }{2-q} \biggr] 
\\ \nonumber \approx 
  \frac{\sT\up_B Q_0 t_{esc}}{\pi R^{\p 2}_B}\ \frac{\g_c^{3-q}}{3-q}\ ,
\end{eqnarray}
while in the fast-cooling regime, 
\begin{eqnarray}
\up_{sy,tot} = \frac{\sT\up_B Q_0 t_{esc}\g_c\g_1^{-q-1}}{\pi R^{\p 2}_B}
\biggr[ \g_1^2(\g_1-\g_c) + 
\\ \nonumber
\g_1^{q+1}\frac{ \g_2^{-q+2}-\g_1^{-q+2}}{2-q} \biggr] 
\\ \nonumber \approx 
\frac{\sT\up_B Q_0 t_{esc}\g_c}{\pi R^{\p 2}_B}\g_1^{2-q} \ .
\end{eqnarray}
 
For a given set of parameters, the nonlinear nature of $\up_{sy}$
means that the above equations do not have a simple closed form
solution for $N_e(\g)$.  The nonlinear effects of SSC cooling have
been explored in detail by \citet{schlickeiser09, schlickeiser10}; and
\citet{zacharias10,zacharias12}.  We solve for $N_e(\g)$ numerically.

The observed isotropic synchrotron luminosity in the frame of the AGN 
in the $\delta$-approximation \citep[e.g.,][]{dermer02,dermer09_book} is
\begin{eqnarray}
L_\e^{sy} = \e L_{sy}(\e) = \frac{2\dD^4}{3} c\sT \up_B \g^3 N_e(\g)\ ,
\end{eqnarray}
where
\begin{eqnarray}
\g = \sqrt{\frac{\e}{\dD\e_B}}\ ,
\end{eqnarray}
$\e$ is the dimensionless emitted photon energy, $\e_B = B/B_{cr}$,
and $B_{cr} = 4.414\times10^{13}$\ G.  For $\g<\min(\g_1,\g_c)$,
$L_{\e}^{sy} \propto \e^{4/3}$, assuming it remains optically thin.
The synchrotron peak will occur at $\g=\g_c$ in the slow-cooling
regime, and at $\g=\g_1$ in the fast cooling regime, so the peak
synchrotron luminosity will be
 
\begin{equation}
\label{Lpk}
L_{pk}^{sy} = L_{\e_{pk}}^{sy} = \frac{2\dD^4}{3} c\sT \up_B Q_0 t_{esc} 
\left\{ \begin{array}{ll}
\g_c^{3-q} & \g_1 < \g_c \\
\g_c\g_1^{2-q} & \g_c < \g_1 
  \end{array}
\right. \ .
\end{equation}
 
This peak will occur at
\begin{equation}
\label{epk}
\e_{pk} = \frac{h\nu^{sy}_{pk}}{m_ec^2} = \frac{\nu^{sy}_{pk}}{1.23\times10^{20}\ \Hz} = 
\dD\e_B \left\{ \begin{array}{ll}
\g_c^2 & \g_1 < \g_c \\
\g_1^2 & \g_c < \g_1 
  \end{array}
\right. \ .
\end{equation}

The synchrotron luminosity in the frame of the blob, $L_{pk}^{\p sy}$ is given 
by Equation (\ref{Lpk}) with $\dD=1$.  From this one can find the 
peak synchrotron energy density,  
\begin{equation}
\label{usypk}
\up_{sy,pk} = \frac{R^{\p}_b}{c} \frac{L_{pk}^{\p sy}}{4\pi R^{\p 3}_b/3}
 = \frac{\up_B \sT}{2\pi R_b^{\p 2}}\ Q_0 t_{esc}\ 
\left\{ \begin{array}{ll}
\g_c^{3-q} & \g_1<\g_c \\
\g_c\g_1^{2-q} & \g_c<\g_1
  \end{array}
\right. \ .
\end{equation}
The SSC luminosity at the peak in the Thomson regime 
can be approximated by \citep{finke08_SSC}
\begin{equation}
L_{pk}^{SSC} = \frac{2\dD^4}{3}c\sT \up_{sy,pk}Q_0 t_{esc}\ 
\left\{ \begin{array}{ll}
\g_c^{3-q} & \g_1 < \g_c \\
\g_c\g_1^{2-q} & \g_c < \g_1 
  \end{array}
\right. \ ,
\end{equation}
or, using Equation (\ref{usypk}),
\begin{equation}
L_{pk}^{SSC} = \frac{\dD^4}{3\pi R_b^{\p 2}} c \sT^2 \up_B (Q_0 t_{esc})^2\ 
\left\{ \begin{array}{ll}
\g_c^{6-2q} & \g_1 < \g_c \\
\g_c^2\g_1^{4-2q} & \g_c < \g_1 
  \end{array}
\right. \ .
\end{equation}
 
The peak luminosity from Thomson-scattering an external isotropic
radiation field in the $\delta$-approximation is \citep[e.g.,][]{dermer02}
 
\begin{equation}
L_{pk}^{EC} = \dD^6 c\sT u_{ext}Q_0 t_{esc}\ 
\left\{ \begin{array}{ll}
\g_c^{3-q} & \g_1 < \g_c \\
\g_c\g_1^{2-q} & \g_c < \g_1 
  \end{array}
\right. \ .
\end{equation}

The Compton dominance ($A_C$) is given by the ratio of the peak
Compton-scattered component to the peak of the synchrotron component,
\begin{eqnarray}
\label{CDdef}
A_C \equiv \frac{\max[L^{EC}_{pk},\ L^{SSC}_{pk}]}{L^{sy}_{pk}} \approx 
\frac{\max[ \dD^2 u_{ext},\ \up_{sy,pk}]}{\up_B}\ ,
\end{eqnarray}
where we have ignored a bolometric correction term $\sim 1$.  Note
that SSC emission has the same beaming pattern as synchrotron, and
thus $L^{SSC}_{pk}/L^{sy}_{pk}$ does not depend on the viewing angle;
however, $L^{EC}_{pk}$ does not have the same beaming pattern as
synchrotron and SSC, and so $L^{EC}_{pk}/L^{sy}_{pk}$ is dependent of the
viewing angle through $\dD$ \citep{dermer95,georgan01}.

\subsection{Results}
\label{modelresults}

We would like to use the model described in Section \ref{blazarmodel}
to phenomenologically reproduce the blazar properties shown in Figures
\ref{fossati} and \ref{CD}.  Blazars are necessarily not observed at
the same angle, and we would like to take this into account.  By
contrast, in previous attempts to explain the ``blazar sequence''
\citep[e.g.,][]{ghisellini98,boett02_seq,ghisellini10}, authors
usually assume all objects are viewed at an angle such that $\delta_D
= \Gamma$.  Further, we would like to do this by varying the least
number of parameters possible between objects.  Finally, we note that
we attempt to reproduce Figures \ref{fossati} and \ref{CD} directly,
something previous authors have not done.  We are attempting to
reproduce the population as a whole, and not individual blazars.

\begin{figure}
\vspace{5.0mm}
\epsscale{1.0}
\plotone{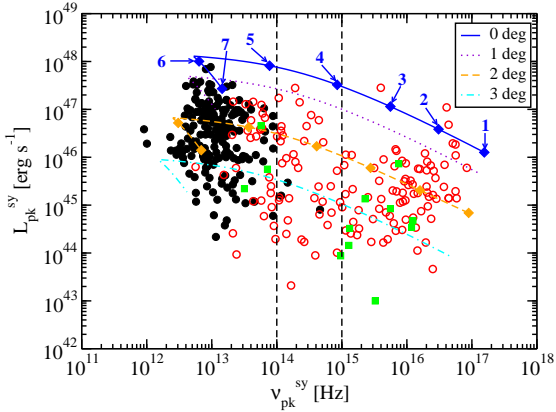}
\caption{Same as Figure \ref{fossati}, with curves showing our model
plotted at various angles, shown in the legend.  Along the curves only
$B$ and $u_{ext}$ are varied.  Model parameters are found in Table
\ref{table_params}.  Diamond symbols show the location of the model
SEDs plotted in Figures \ref{SED} and \ref{SED_2deg}, with the
numbers indicating the location of the SEDs in Figure \ref{SED}.} 
\label{fossati_theory}
\vspace{2.2mm}
\end{figure}

\begin{figure}
\vspace{3.0mm}
\epsscale{1.0}
\plotone{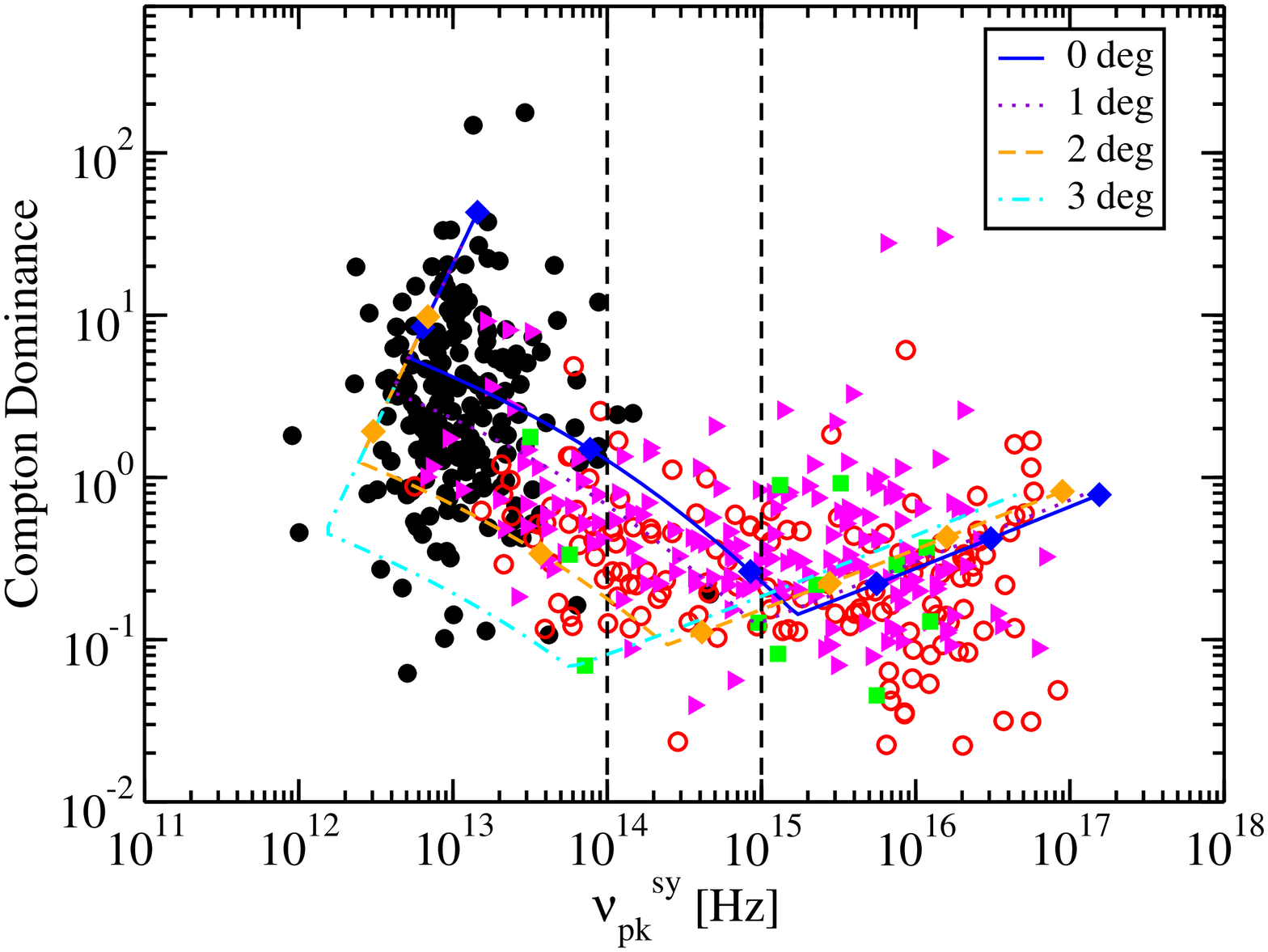}
\caption{Same as Figure \ref{CD} with curves showing our model
plotted at various angles, shown in the legend.  Along the curves only
$B$ and $u_{ext}$ are varied.  Model parameters are found in Table
\ref{table_params}.  Diamond symbols show the location of the model
SEDs plotted in Figures \ref{SED} and \ref{SED_2deg}.}  
\label{CD_theory}
\vspace{2.2mm}
\end{figure}

Our modeling results are shown in Figures \ref{fossati_theory} and
\ref{CD_theory}.  The curves show the sources at at constant angle,
according to the model, so that along the curve only $B$ and $u_{ext}$
vary according to Equation \ref{B_uext_eqn}.  The model parameters are
shown in Table \ref{table_params}, and are fairly close to the values
used by \citet{boett02_seq}.  The sources with low $B$ and low
$u_{ext}$ have high $\nu_{pk}^{sy}$, due to the small amount of
cooling.  As $B$ and $u_{ext}$ increase, the cooling increases, and
hence $\g_c$ decreases.  Since $\nu_{pk}^{sy} \propto \e_{pk}\propto
\g_c^2 B$ (Equation [\ref{epk}]), and $\g_c$ decreases faster than $B$
increases, $\nu_{pk}^{sy}$ will decrease.  If the cooling is great
enough, and $\g_c<\g_1$, the curves enter the fast cooling regime, and
$\nu_{pk}^{sy}$ is associated with $\g_1$ instead of $\g_c$ (again
Equation [\ref{epk}]).  The transition between the fast and slow
cooling regimes leads to the sharp break at $\la 10^{13}$\ Hz (the
exact location depends on the angle) seen in the curves in Figures
\ref{fossati_theory} and \ref{CD_theory}.  In Figure \ref{CD_theory},
there is also a break in the curves between $\approx 10^{14}$ and
$\approx 10^{15}$\ Hz caused by the transition between SSC and EC.
Note that for SSC, $A_C$ does not depend on angle, while for EC, it
does.  The curves reproduce almost all of the objects in Figures
\ref{fossati_theory} and \ref{CD_theory}, although there are
objects with $A_C\la0.1$ and $\nu_{pk}^{sy}>10^{15}$\ Hz that are not
reproduced.  A possible explanation is if there exist structures
within the jet, so that as one views a jet more and more off axis, one
views a different region of the jet \citep{meyer11}.  This would only
work if $B$, $Q_0$, and/or $R_b^{\p}$ also varied not just $\delta_D$,
since $A_C$ is independent of $\delta_D$ in the case of SSC.

According to this simple model, all sources have $\Gamma=30$.  Of
course, in reality, sources will not have jets moving at the same
speed, as shown by very long baseline interferometry \citep[VLBI;
e.g.,][]{jorstad05,lister09,piner08,piner10}, which also typically
show slower values for $\Gamma$.  However, the main emitting regions
could be on smaller size scales than is possible to resolve with VLBI,
and so may have different speeds, although note that much lower
\citep[e.g.,][]{piner08} and much higher \citep[e.g.,][]{marscher10}
Lorentz factors have been observed.  It has also been suggested that
the jet of a single source may have velocity gradients
\citep[e.g.,][]{chiaberge00, stawarz02,georgan03, ghisellini05,
meyer11}.  However, we assume a gradient in $\Gamma$ is not necessary
to reproduce the general trends \citep[cf.][]{meyer11}, and this
value may represent an average value for the main emitting region.
The magnetic field strength values, which span from $B=0.050$\ G to
$B=8.6$\ G, are consistent with those found from spectral modeling of
blazars \citep[e.g.,][]{ghisellini98,ghisellini10}.  These values are
also consistent in that the modeling generally shows higher $B$ for
FRSQs than BL Lacs.  

Blazar SED modeling also indicates that the $\g$-ray emission from
FSRQs is likely from Compton scattering of an external radiation
source, while for HSP BL Lacs SSC is able to provide a good fit to the
$\g$-ray emission \citep[e.g.,][]{ghisellini98,ghisellini10}.  A
correlation between $A_C$ and core dominance, a proxy for $\theta$,
found in FSRQs, is also evidence that EC dominates in these sources
\citep{meyer12}.  BL Lacs tend to have less prominent ``blue bumps''
from accretion disks and weaker broad lines, two of the leading
contenders for the seed photon source for external Compton scattering.
The presumed parent population of BL Lacs, FR~I radio galaxies, also
seem less likely to have dust tori \citep{donato04}, another leading
contender for the external photon source.  {\em Wide Field Infrared
Survey Explorer} observations of BL Lacs do not show any evidence for
dust tori \citep{plotkin11}.  So whatever the source is, it seems
reasonable to assume it is greater for more FSRQs, and weaker for BL
Lacs.  This is similar to \citet{boett02_seq} but in contrast to
Ghisellini et al., who use a ``binary'' $u_{ext}$ where the external
radiation field is either ``on'' or ``off'' above and below a certain
accretion rate \citep[e.g.,][]{ghisellini08}.  They justify this from
reverberation mapping campaigns, which show the disk luminosity is
proportional to the square of the BLR distance from the disk
\citep[e.g.,][]{bentz06}, which would result in the same energy
density observed, as long as the emitting region is inside the BLR,
and the same fraction of disk emission is reprocessed by the BLR.
However, there is no guarantee that this fraction is the same in all
sources, or that if the external radiation source is a dust torus that
it follows the same relation as the BLR.  It seems more likely that
$u_{ext}$ can have a range of values, rather than just one.  Whether
the external photon source is a dust torus or BLR, its luminosity
should be less than the Eddington limit for a black hole with mass
$M=10^9 M_{\odot}$, or $L_{Edd}=1.3\times10^{47} M_9\ \erg\ \s^{-1}$
where $M_9 = M / (10^9 M_{\odot})$.  If the radius of the external
source is $R_{ext}=10^{18}\ \cm \approx 1/3\ \pc$, then the maximum
energy density will be
\begin{eqnarray}
u_{ext} = \frac{L_{Edd}}{4\pi c R_{ext}^2} = 0.35\ \erg\ \cm^{-3}\ M_9 R_{18}^{-2}
\end{eqnarray}
where $R_{18} = R_{ext} / (10^{18}\ \cm)$.  It is unlikely though that
the external photon source will be radiating all of the black hole's
accretion luminosity, so $u_{ext}$ should be lower than this by at least a
factor 10.  We have thus chosen our parameters so that the maximum
external energy density is $u_{ext}=0.035\ \erg\ \cm^{-3}$.

We have also computed the jet powers in Poynting flux and
electrons, respectively, by
\begin{eqnarray}
P_{j,B} = 2\pi R_b\p \G^2 c\beta \up_B
\end{eqnarray}
and
\begin{eqnarray}
P_{j,e} = 2\pi R_b\p \G^2 c\beta \up_e\ 
\end{eqnarray}
\citep{celotti93,celotti07,finke08_SSC}.  These results are also shown
in Table \ref{table_params_individ}.  The jet powers are consistent
with those found previously by other authors for spectral modeling
\citep[e.g.,][]{ghisellini09,ghisellini10}.  For sources with high
$B$, naturally $P_{j,B}$ is higher.  In the slow cooling regime,
$P_{j,e}$ is nearly independent of $B$ and $u_{ext}$, decreasing with
increasing $B$ and $u_{ext}$ only slightly.  In the fast cooling
regime (Equation [\ref{fast_cool_elec}]), $N_e(\g)$, and hence $u_e$,
is directly proportional to $\g_c$, so that as $B$ and $u_{ext}$
increase, $\g_c$ decreases, leading to lower $u_e$.

\begin{deluxetable}{lcc}
\tabletypesize{\scriptsize}
\tablecaption{
Parameters (described in the text) for theoretical curves in 
Figures \ref{fossati_theory} and \ref{CD_theory}.
}
\tablewidth{0pt}
\tablehead{
\colhead{Parameter} &
\colhead{Value} 
}
\startdata
$\g_1$ & $10^2$ \\
$\g_2$ & $10^6$ \\
$R_b$ & $10^{16}\ \cm$ \\
$\eta$ & $1.0$ \\
$L_{inj}$ & $2.0\times10^{42}\ \erg\ \s^{-1}$ \\
$Q_0$ & $6\times10^{48}\ \s^{-1}$ \\
$q$ & 2.5 \\
$u_0$ & $4.0\times10^{-11}\ \erg\ \cm^{-3}$ \\
$\G$ & $30$ \\
$B_0$ & $0.05$\ G \\
\enddata
\label{table_params}
\end{deluxetable}

\begin{figure}
\vspace{3.0mm}
\epsscale{0.91}
\plotone{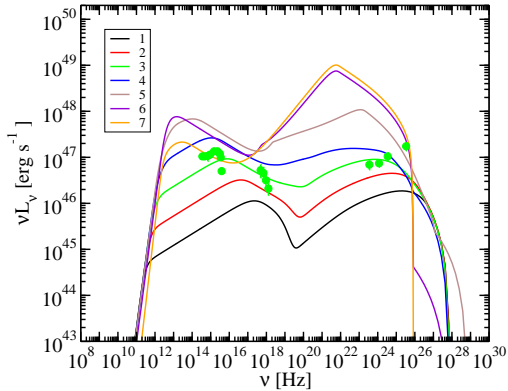}
\caption{SEDs for the sequence seen in Figures
\ref{fossati_theory} and \ref{CD_theory} as observed at $\theta=0$.  
The numbers correspond to the numbered diamonds in Figure \ref{fossati_theory}.
The circles are the SED data for CRATES 0630$-$2406 from
\citet{padovani12}. }
\label{SED}
\vspace{2.2mm}
\end{figure}

\begin{figure}
\vspace{2.2mm}
\epsscale{0.91}
\plotone{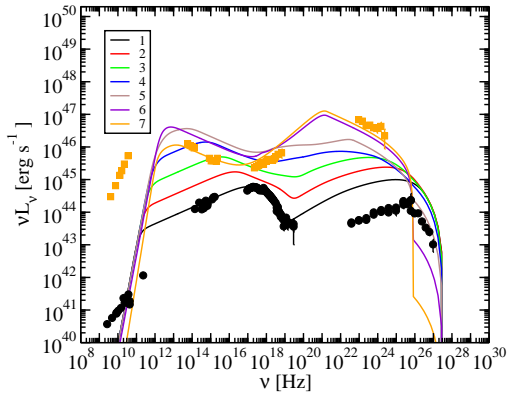}
\caption{SEDs for the sequence (diamonds) seen in Figures
\ref{fossati_theory} and \ref{CD_theory} as observed at
$\theta=2\arcdeg$.  The circles are the SED data for Mrk~421 from
\citet{abdo11_mrk421}, and the squares are the SED data for 3C~279
from \citet{hayashida12}.  }
\label{SED_2deg}
\vspace{2.2mm}
\end{figure}

The SEDs of several blazars along the sequence, seen at an angle to
the line of sight $\theta=0$ and $\theta=2\arcdeg$ can be seen in
Figures \ref{SED} and \ref{SED_2deg}, respectively.  These SEDs
correspond to the diamonds in Figures \ref{fossati_theory} and
\ref{CD_theory}.  The parameters for these curves can be found in
Table \ref{table_params_individ}.  The external radiation field is
assumed to be isotropic in the galaxy's frame and monochromatic, with
dimensionless seed photon energy $\e_0=5\times10^{-7}$ in the galaxy's
frame, which is around what one would expect for the peak emission
from a dust torus with temperature $T_{dust}\sim10^3\ \Kelvin$.  The
calculations were performed using the exact synchrotron emissivity and
the full Compton cross section, accurate in the Thomson through
Klein-Nishina regimes.  The details of these calculations can be found
in \citet{finke08_SSC} and \citet{dermer09}.  Since exact expressions
are used, the parameters $L_{pk}^{sy}$, $\nu_{pk}^{sy}$, and $A_C$
differ slightly from the values found in Figures \ref{fossati_theory}
and \ref{CD_theory}.  Note that synchrotron-self absorption is
included in these curves as well, and that it is more apparent for the
higher power blazars, where the magnetic field is larger and the
emitting region becomes self-absorbed at higher frequencies.  A
contribution from an underlying accretion disk is not included,
although this has been shown to dominate the optical continuum for
FSRQs and some BL Lacs.  These curves appear similar to SEDs observed
from blazars.  Similar to \citet{boett02}, we do not attempt to
reproduce any individual blazar.  However, to demonstrate the
similarity to actual blazar SEDs, we plot data for several blazars on
these Figures as well.  These include the CRATES 0630$-$2406 from
\citet{padovani12} in Figure \ref{SED}, and the Mrk~421 SED from
\citet{abdo11_mrk421} and the low state 3C~279 from
\citet{hayashida12}, in Figure \ref{SED_2deg}.  The symbols have the
same color as the curve which is the closest match.  The curves are
not a perfect fit to the SED data, although they are a reasonable
representation.

The scenario described here predicts that a large fraction of FSRQs
are emitting in the fast-cooling regime.  In this regime
($\g_c<\g_1$), one expects the electron index below the peak
(associated with $\g_1$) to be $p=2$, as shown in Equation
(\ref{fast_cool_elec}).  For FSRQs, Compton scattering by these
electrons should make hard X-rays ($\sim$ 10 -- 100 keV) and the
spectral index from Thomson scattering would then be
$\G_i=(p+1)/2=1.5$, assuming one is observing reasonably far below the
break.  In the {\em Swift}-BAT spectra from 36 months of observations
\citep{ajello09}, almost every FSRQ has an X-ray spectral index
consistent with $\G_i=1.5$ within the error bars, and the mean for the
sample is $1.6\pm0.3$.  The BeppoSAX six year catalog \citep{donato05}
has a large number of sources are near $\G_i\approx 1.5$ with a mean
spectral index of $1.59\pm 0.05$, although not all sources in that
catalog are consistent with $\G_i=1.5$.  The diversity of hard X-ray
spectral indices could be explained by the location of the Compton
component within the BAT or BeppoSAX bandpass.  Softer spectra
($\G_i>1.5$) could be those with their Compton peak being in or near
the BAT or BeppoSAX bandpass, or by a contribution from SSC emission.
Harder spectra ($\G_i<1.5$) could be caused by the photons scattered
by electrons below $\g_c$ being in the hard X-ray bandpass.

\begin{deluxetable*}{lccccccc}
\tabletypesize{\scriptsize}
\tablecaption{
Parameters (described in the text) for curves shown in 
Figures \ref{SED} and \ref{SED_2deg}.
}
\tablewidth{0pt}
\tablehead{
\colhead{Parameter} &
\colhead{1} &
\colhead{2} &
\colhead{3} &
\colhead{4} &
\colhead{5} &
\colhead{6} &
\colhead{7} 
}
\startdata
$B$ [G] & $0.050$ & $0.12$ & $0.28$ & $0.67$ & $1.6$ & $3.8$ & $8.6$ \\
$u_{ext}$ [$\erg\ \cm^{-3}$] & $4\times10^{-11}$ & $1.3\times10^{-9}$ & $4.1\times10^{-8}$ & $1.3\times10^{-6}$ & $4.2\times10^{-5}$ & $1.4\times10^{-3}$ & $3.5\times10^{-2}$ \\
$\g_c$ & $1.4\times10^{5}$ & $3.9\times10^4$ & $1.1\times10^4$ & $2.7\times10^{3}$ & $5.4\times10^2$ & $50$ & $2.7$ \\
$P_{j,B}$ [$\erg\ \s^{-1}$] & $1.7\times10^{42}$ & $9.6\times10^{42}$ & $5.4\times10^{43}$ & $3.1\times10^{44}$ & $1.7\times10^{45}$ & $9.8\times10^{45}$ & $5.0\times10^{46}$ \\
$P_{j,e}$ [$\erg\ \s^{-1}$] & $1.3\times10^{45}$ & $1.3\times10^{45}$ & $1.2\times10^{45}$ & $1.2\times10^{45}$ & $9.5\times10^{44}$ & $4.5\times10^{44}$ & $7.6\times10^{43}$   \\
\enddata
\label{table_params_individ}
\end{deluxetable*}
\vspace{2.2mm}

\section{Summary and Discussion}
\label{discussion}

In Section \ref{2lac_sequence} we have found anti-correlations in the
2LAC that were found previously by many authors in other samples.  The
plot and correlation with $A_C$ in particular is novel, since the 2LAC
allows for a larger sample of the Compton component than any previous
catalog.  We include on this plot for the first time a large number of
blazars without known $z$, and showed that the anti-correlation
with $\nu_{pk}^{sy}$ is not a relic of ignoring blazars which lack
spectroscopic redshifts.  {\em This relationship seems to have a
physical origin.}

Both \citet{meyer12} and \citet{ghisellini12_seq} plot the LAT
$\gamma$-ray luminosity ($L_\gamma$) versus the LAT spectral index
($\Gamma_\g$; a proxy for $\nu_{pk}^C$), and showed that the four
sources from \citet{padovani12} with high $\nu_{pk}^{sy}$, high
$L_{pk}^{sy}$ do not have excessively large $L_{\g}$ or excessively
small $\G_\g$, and are consistent with other LAT blazars.  However, it
is possible that the redshifts of many BL Lacs in the future could be
measured, and they could be found with high $L_\gamma$ and low
$\G_\g$.  So it is not possible to assess this selection effect
with a plot of $L_\gamma$ versus $\Gamma_\g$.

In Section \ref{theory} we developed a simple model to explain the
plots and anti-correlations from Section \ref{2lac_sequence}.  This
shape is explained by the increased external radiation field energy
density and magnetic field strength, where $\nu_{pk}^{sy}$ is
associated with either $\g_c$ (the Lorentz factor of the break in the
electron spectrum due to cooling) for sources in the slow-cooling
regime, and $\g_1$ (the lowest Lorentz factor of the injected
electrons) for sources in the fast-cooling regime.

This theory is quite simple, and neglects numerous important effects.
Perhaps most importantly, blazars are highly variable, and we treat
them with a steady-state solution to the electron continuity equation.
In any study of a large population of blazars, however, this is almost
unavoidable.  In developing this theory, approximate expressions were
used, neglecting the exact synchrotron emissivity and Compton cross
section, especially Klein-Nishina effects.  However, a comparison
between the predictions and exact calculations (Figure \ref{SED})
shows that these approximations seem reasonable.

Despite its simplicity, this model makes a number of predictions.  It
predicts that the single most important parameter in determining the
luminosity ($L_{pk}^{sy}$) of blazars is $\theta$, mostly independent
of the frequency of their synchrotron peak (Figure
\ref{fossati_theory}).  This also means that as fainter sources
are found, these sources will not have excessively large values for
$A_C$, since large values for this quantity are due to EC, and $A_C$
is strongly dependent on $\theta$ if EC is dominant.  There should not
be any sources found with $L_{pk}^{sy} \la 10^{45}\ \erg\ \s^{-1}$ and
$A_C \ga$ a few.  This model also predicts that no ``blue FSRQs'',
that is, no HSPs with significant BLR or dust torus luminosity will be
found.  It predicts that any source with high $A_C$ ($\ga$\ a few)
will be an LSP and it will be in the fast cooling regime, so that the
spectral index of the EC emission below the $\g$-ray peak will be
$\G_i\sim 1.5$.  For most sources, this will be in the soft $\g$-rays
($\la 10$\ MeV) down to the hard X-ray regime, perhaps as low as 10
keV in some cases.

This model predicts that if the BL Lacs sources become more
redshift-complete, more sources with high $L_{pk}^{sy}$ and
$\nu_{pk}^{sy}$ (``bright HSPs'') will be found, although these
sources will not be as bright as the brightest LSPs (Figure
\ref{fossati_theory}).  These sources will have their jets highly
aligned with our line of sight.  Note that our scenario differs
somewhat from the one of \citet{ghisellini12_seq}.  They also predict
that bright HSP sources will be found, but in their case, these
sources are ``blue FSRQs'', with the primary emitting region found
outside the BLR, avoiding an EC component and a large $A_C$, since the
$\g$-ray emission will be due only to SSC.  Their scenario and ours
would produce essentially identical SEDs for bright HSPs, so
distinguishing them is not possible based on SEDs alone.  If it could
be found that these sources have their jets highly aligned, that they
do not have significant BLRs or dust tori, or that the BLRs are not
the seed photon sources for EC in ``red'' (i.e., LSP) FSRQs, it
would favor our scenario.  All of these things are difficult to
determine observationally, however.  The nonthermal synchrotron makes
broad emission lines or a dust component nearly impossible to observe,
although jet alignment may be possible to determine with VLBI
observations \citep[e.g.,][]{jorstad05}.

Our model can also be contrasted with the Monte Carlo simulations of
\citet{giommi12_selection}.  In their simulations, they randomly draw
for each blazar several properties including $\gamma_{pk}$ (the peak
of the electron distribution) $\delta_D$, and the strength of the
broad emission lines, all independent of each other.  Although
they do not explore the Compton dominance in their paper, if these
properties are in reality independent of each other, and if the BLR is
the seed photon source for Compton scattering, the scenario of
\citet{giommi12_selection} would not produce a correlation between
$A_C$ and $\nu_{pk}^{sy}$, and would produce objects with high
$A_C$ and high $\nu_{pk}^{sy}$.  Since no such objects are found
(Figure \ref{CD}) and a correlation between $A_C$ and $\nu_{pk}^{sy}$
is found (Table \ref{table_correlate}), their scenario does not seem
to be in agreement with observations.

Our simple model assumes the primary emitting region has a single
Lorentz factor.  Based on optical \citep{chiaberge00} and $\g$-ray
\citep{abdo10_cena,abdo10_misalign} observations of radio galaxies, it
seems highly likely that jets are stratified in speed either
perpendicular \citep{stawarz02,ghisellini05} or parallel to the jet's
direction of motion \citep{georgan03}.  Indeed, \citet{meyer11} have
an explanation for an ``L'' shape in a plot of $L_{pk}^{sy}$
versus $\nu_{pk}^{sy}$, based on sources with varying Lorentz factors
within a source.  In our study we neglect radio galaxies and
stratified jets.  However, modeling of radio galaxies
\citep{chiaberge01,abdo09_m87,abdo10_sed,migliori11} indicates that
these types of structures almost certainly exist.  Whether they can
explain the properties of blazars as well as radio galaxies remains to
be seen.

Another possible way to decipher the correct model could involve
estimating the power injected into the jet.  This power can be related
to the power needed to create a cavity in the hot X-ray emitting
ICM surrounding radio galaxies
\citep{birzan04,birzan08,cavagnolo10}.  This power seems to be
correlated with the extended lobe's radio power, and \citet{meyer11}
have used the radio power as a proxy for jet kinetic power.  This
angle-independent measure of the jet kinetic power may provide a way
of distinguishing viewing angle effects from intrinsic power effects.

As a redshift-independent quantity, the Compton dominance is a useful
tool for exploring blazar properties, including the large number of BL
Lacs without known redshifts.  The large new blazar catalog from
the {\em Fermi}-LAT is able to characterize the Compton component for
a larger number of objects than previously possible, making it
valuable for determining its relationship to the blazar sequence.

\acknowledgements 

We are grateful to C.D.\ Dermer, M.\ Georganopoulos, M.\ Lister, and
K.\ Wood for useful discussions on the blazar sequence, to C.D. Dermer
and A.\ Wehrle who gave helpful comments on a previous version of this
manuscript, to M.\ Hayashida for the SED of 3C~279, and to the
anonymous referee for very useful comments.  This work was partially
supported by {\em Fermi} GI Grant NNH09ZDA001N.




\bibliographystyle{apj}
\bibliography{3c454.3_ref,EBL_ref,references,mypapers_ref,blazar_ref,sequence_ref,SSC_ref}

\end{document}